\documentclass[12pt]{article}

\usepackage{graphicx}
\usepackage{float}
\usepackage{cite}

\def\gtwid{\mathrel{\raise.3ex\hbox{$>$\kern-.75em\lower1ex\hbox{$\sim$}}}}
\def\ltwid{\mathrel{\raise.3ex\hbox{$<$\kern-.75em\lower1ex\hbox{$\sim$}}}}
\def\square{\kern1pt\vbox{\hrule height 1.2pt\hbox{\vrule width 1.2pt\hskip 3pt
   \vbox{\vskip 6pt}\hskip 3pt\vrule width 0.6pt}\hrule height 0.6pt}\kern1pt}

\begin{document}

\begin{titlepage}

\begin{flushright}
BRX-TH 6648 \\
CALT-TH 2019-008 \\
UFIFT-QG-19-02
\end{flushright}

\vskip .5cm

\begin{center}
{\bf Nonlocal Cosmology II --- Cosmic acceleration without fine tuning or 
dark energy}
\end{center}

\begin{center}
S. Deser$^{1*}$ and R. P. Woodard$^{2*}$
\end{center}

\begin{center}
\it{$^{1}$ California Institute of Technology, Pasadena, CA 91125 and \\
Department of Physics, Brandeis University, Waltham, MA 02254}
\end{center}

\begin{center}
\it{$^{2}$ Department of Physics, University of Florida,\\
Gainesville, FL 32611}
\end{center}

\vspace{0.5cm}

\begin{center}
ABSTRACT
\end{center}
We present an improved version of our original cosmological model to explain
the current phase of cosmological acceleration without resorting to a 
cosmological constant or any other mass scale. Like the original, this 
phenomenological approach is based on an effective quantum gravitational 
action, but now depends on the original nonlocal dimensionless scalar $X = 
\square^{-1} R$ only through $Y = \square^{-1} g^{\mu\nu} X_{,\mu} X_{,\nu}$. 
Both $X$ and $Y$ are quiescent during the radiation-dominated ($R=0$) era, 
both only grow logarithmically during matter dominance, and neither affects 
the propagation of gravitational radiation. However, while $X$ has the same 
sign for gravitationally bound systems as for cosmology, we show that the 
sign of $Y$ differs for the two cases: it is positive for cosmology and 
negative for strongly gravitationally bound systems. We can therefore enforce 
the $\Lambda$CDM expansion history by making a suitable choice of the nonlocal 
distortion function $f(Y)$ for $Y > 0$, while ensuring that there is no 
change in the heavily constrained solar system phenomenology simply by making 
$f$ vanish for $Y < 0$ without discontinuity. The required $f(Y>0)$ is 
determined numerically to have a strikingly simple exponential form.

\begin{flushleft}
\end{flushleft}

\begin{flushleft}
$^{*}$ e-mail: deser@brandeis.edu , woodard@phys.ufl.edu
\end{flushleft}

\end{titlepage}

\section{Introduction}

This work is a continuation, and shares the basis and philosophy, 
of our original cosmological model \cite{Deser:2007jk,Deffayet:2009ca,
Deser:2013uya}. The present cosmological acceleration phase of the universe 
\cite{Abbott:2018xao} is a major, if originally unexpected, feature of late 
time expansion. An explanation not invoking new physics or fine tuning is 
clearly to be preferred; ours was a nonlocal one, based on a
function of the dimensionless scalar $X[g] = \square^{-1} R$. The argument 
was that it represents current effects of the necessarily abundant infrared 
gravitons in the early universe \cite{Woodard:2014iga,Woodard:2018gfj}.

Extensive studies have been made of the theory's cosmological perturbations
\cite{Park:2012cp,Dodelson:2013sma,Park:2016jym,Nersisyan:2017mgj,
Park:2017zls,Amendola:2019fhc}. There have also been studies of future
cosmological evolution \cite{Koivisto:2008xfa}, solar system constraints 
\cite{Koivisto:2008dh}, and the generation of gravitational radiation
\cite{Chu:2018mld}.\footnote{A similar model has also been proposed which is 
based on the dimensionful nonlocal scalar $\square^{-2} R$ 
\cite{Maggiore:2014sia,Maggiore:2016gpx}. Many studies have been made of the 
phenomenology of this model \cite{Dirian:2014ara,Barreira:2014kra,
Dirian:2014xoa,Dirian:2014bma,Dirian:2016puz,Nersisyan:2016hjh}.}

Our original model assumed that $X[g]$ had opposite signs in the cosmological 
($-$) and the (smaller scale) gravitationally bound ($+$) contexts. That
would prevent --- unwanted --- effects in the latter. However, it was 
recently pointed out that $X[g]$ is negative definite \cite{Belgacem:2018wtb}.
We overcome this difficulty by a simple modification: replacing $X[g]$ by the 
(equally nonlocal) $Y[g] \equiv \square^{-1} [g^{\mu\nu} \partial_{\mu} X 
\partial_{\nu} X]$, removes the problem without losing the explanation of 
accelerated expansion: While both $X[g]$ and $Y[g]$ vanish during radiation 
domination ($R = 0$), and only grow slowly thereafter, $Y[g]$ does have 
opposite signs in strongly bound matter ($Y < 0$) and in the large ($Y > 0$); 
so we merely define the nonlocal distortion function $f(Y)$ to vanish for 
$Y < 0$, and have the proper details for $Y > 0$, thus restoring the desired 
behavior throughout.\footnote{It turns out that enforcing the $\Lambda$CDM
requires $f(Y) \rightarrow Y^2 \ln(Y)$ as $Y$ approaches zero from above. 
This means that making $f(Y<0) = 0$ leads to no discontinuity in either $f(Y)$
or $f'(Y)$ at $Y=0$.} An additional, highly desirable property of both the 
original and the new theories is that there is no change in the constrained 
propagation of gravitational radiation \cite{Boran:2017rdn}.

Section 2 defines our model and discusses how it might emerge from fundamental
considerations. It also explains why the new nonlocal scalar $Y[g]$ changes 
sign from cosmological to gravitationally bound systems. Section 3 gives an 
explicit numerical determination of the nonlocal distortion function $f(Y)$ to 
reproduce the $\Lambda$CDM expansion history without a cosmological constant. 
It also derives an amazingly simple exponential fit to $f(Y)$. Section 4 
presents conclusions.

\section{The New Model}

In this section, we define and discuss the improved model. The original problem 
and its remedy are explained. We close with comments on its possible origin 
in a more fundamental setting.

\subsection{Defining the New Model}

Our two nonlocal scalars are
\begin{equation}
X[g] \equiv \frac1{\square} R \; , \; Y[g] \equiv \frac1{\square} 
\Bigl( g^{\mu\nu} \partial_{\mu} X[g] \partial_{\nu} X[g]\Bigr) \; ; \;
\square \equiv \frac1{\sqrt{-g}} \partial_{\mu} \Bigl( \sqrt{-g} g^{\mu\nu} 
\partial_{\nu}\Bigr) \; , \label{XYdef}
\end{equation}
where $\square^{-1}$ is defined by retarded boundary conditions which require
that $X[g](x)$, $Y[g](x)$ and their first derivatives all vanish on the initial 
value surface. Our nonlocal modification is defined by the distortion function 
$f(Y)$,
\begin{equation}
\mathcal{L}_{\rm nonlocal} \equiv \frac{1}{16 \pi G} \, R \Bigl[1 + 
f\Bigl(Y[g]\Bigr) \Bigr] \sqrt{-g} \; . \label{Lnonlocal}
\end{equation}
Just as the original model could be localized through the introduction of two 
auxiliary scalar fields \cite{Nojiri:2007uq}, the new model requires four 
auxiliaries,\footnote{We shall abuse the notation slightly by using the same
symbols $X$ and $Y$ for auxiliary scalars in the localized model (\ref{Llocal}) 
as for their retarded solutions (\ref{XYdef}).} 
\begin{equation}
\mathcal{L}_{\rm local} \equiv \frac{\sqrt{-g}}{16 \pi G} \Biggl[ R \Bigl(1 + U 
+ f(Y)\Bigr) + \Bigl( \partial_{\mu} X \partial_{\nu} U + \partial_{\mu} Y 
\partial_{\nu} V + V \partial_{\mu} X \partial_{\nu} X\Bigr) g^{\mu\nu} \Biggr] . 
\label{Llocal}
\end{equation}
It is important to bear in mind that the auxiliary scalars do {\it not} have 
arbitrary initial value data, which would result in new degrees of freedom, half 
being ghosts \cite{Deser:2013uya,Park:2019btx}. Instead, all obey retarded 
boundary conditions, hence introduce no excitations. 

The $U$, $V$ Lagrange multipliers, whose variations lead to the equations which, 
with retarded boundary conditions, define $X[g]$ and $Y[g]$,
\begin{eqnarray}
\frac{16 \pi G}{\sqrt{-g}} \frac{\delta S}{\delta U} & = & -\square X + R = 0 
\qquad \Longrightarrow \qquad X[g] = \frac1{\square} R \; , \label{Ueqn} \\
\frac{16 \pi G}{\sqrt{-g}} \frac{\delta S}{\delta V} & = & -\square Y +
g^{\mu\nu} X_{,\mu} X_{,\nu} = 0 \quad \Longrightarrow \quad Y[g] = \frac1{\square} 
\Biggl[ g^{\mu\nu} X_{,\mu} X_{,\nu} \Biggr] \; . \qquad \label{Veqn}
\end{eqnarray}
Variation with respect to $X$ and $Y$ leads to similar equations for $U$ and $V$, 
which are also solved with retarded boundary conditions,
\begin{eqnarray}
\frac{16 \pi G}{\sqrt{-g}} \frac{\delta S}{\delta X} & \!\!=\!\! & -\square U - 
2 D_{\mu} (V D^{\mu} X) = 0 \;\; \Longrightarrow \;\; U[g] = -\frac2{\square} 
D_{\mu} (V D^{\mu} X) \; , \quad \label{Xeqn} \\
\frac{16 \pi G}{\sqrt{-g}} \frac{\delta S}{\delta Y} & \!\!=\!\! & -\square V + 
R f'(Y) = 0 \quad \Longrightarrow \quad V[g] = \frac1{\square} R f'(Y) \; . 
\label{Yeqn}
\end{eqnarray}
Note that all four auxiliary scalars propagate along
the characteristic curves of the scalar d`Alembertian $\square$, so the sound
speed should agree with the speed of light, alleviating the problems which can
occur in some modified gravity theories \cite{Sawicki:2015zya}. The gravitational 
field equations are,
\begin{eqnarray}
\lefteqn{ \Bigl(G_{\mu\nu} + g_{\mu\nu} \square - D_{\mu}
D_{\nu}\Bigr) \Bigl(1 + U + f(Y) \Bigr) + \partial_{(\mu} X \partial_{\nu )} U + 
\partial_{(\mu} Y \partial_{\nu )} V } \nonumber \\
& & \hspace{-0.2cm} + V \partial{_\mu} X \partial_{\nu} X - \frac12 g_{\mu\nu}
g^{\rho\sigma} \Bigl( \partial_{\rho} X \partial_{\sigma} U \!+\! \partial_{\rho} Y 
\partial_{\sigma} V \!+\! V \partial_{\rho} X \partial_{\sigma} X \Bigr) = 8\pi G 
T_{\mu\nu} \; ; \quad \label{metriceqn}
\end{eqnarray}
here parenthesized indices are symmetrized and $T_{\mu\nu}$ is the matter 
stress-energy tensor {\it without} dark energy.

\subsection{The signs of $Y$}

To determine the sign of $Y$ in regions of bound matter, we assume the metric there 
to be (quasi-) static, i.e., time-independent and diagonal; it could more generally 
be stationary, $g_{0i} \neq 0$, but the same considerations should still hold with a 
bit more matrix detail. Then $g^{\mu\nu} X_{,\mu} X_{,\nu} \rightarrow g^{ij} X_{,i} 
X_{,j}$ is positive in our ($-+++$) convention. We argue next that, instead, 
$\square^{-1}$ is negative so that there $Y < 0$ also . Recall that in flat space, 
$\square^{-1}$ acting on a time independent source reduces, upon time-integration, to 
$\nabla^{-2}$. But our $\square$ has the flat-space "$-\partial_t^2$" form, namely $-g^{00} 
[-g_{00} \nabla^2 -\partial_t^2]$, except for the overall $-g^{00}$ and the (irrelevant) 
metric dependence of our $\nabla^2$; there is also a (strictly positive) factor $\sqrt{-g} 
= \sqrt{\mbox{}^3g} \sqrt{-g_{00}}$ upstairs. Thus, after time integration, our net 
inverse "Laplacian" is $\frac1{\nabla^2}$ (since $g^{00} g_{00} = 1$), a negative quantity 
when operating on the positive $g^{ij} X_{,i} X_{,j}$. In section 3 we show, by explicit
computation, that $Y$ is positive in the purely time dependent cosmological region.

Another, global, way of understanding the signs of both $X$ and $Y$ is by taking the flat
space limit. The retarded Green's function $G_{\rm ret}[g](x;x')$ which implements 
$\square^{-1}$ reduces, when $g_{\mu\nu} \rightarrow \eta_{\mu\nu}$, to the usual flat
\begin{equation}
G_{\rm ret}[\eta](x;x') = -\frac{\delta(t \!-\! t' - \frac1{c} \Vert \vec{x} \!-\! \vec{x}'
\Vert)}{4\pi \Vert \vec{x} \!-\! \vec{x}' \Vert} \; . \label{flatGF}
\end{equation}
This simple form makes it easy to derive explicit expressions for $X$ and $Y$. For example,
if the Ricci scalar is a positive constant and the initial value surface is at $t = 0$,
\begin{equation}
R(x) = \frac1{\ell^2} \qquad \Longrightarrow \qquad X(x) = -\frac{c^2 t^2}{2 \ell^2}
\quad , \quad Y(x) = +\frac{c^4 t^4}{12 \ell^4} \; . \label{flatXY}
\end{equation}
This situation in which the time dependence of $X$ dominates is relevant to cosmology. 
On the other hand, suppose the Ricci scalar is a positive constant within a sphere of 
radius $\ell$, and we consider some time $t$ much larger than either $\ell$ or $\Vert 
\vec{x} \Vert \equiv r$,
\begin{equation}
R(x) = \frac{\theta(\ell \!-\! \Vert \vec{x} \Vert)}{\ell^2} \; \Longrightarrow \; 
X(x) = -\frac13 \Bigl[ \frac32 \!-\! \frac{r^2}{2 \ell^2}\Bigr] \theta(\ell \!-\! r)
-\frac{\ell}{3 r} \theta(r \!-\! \ell) \; . \label{Xflat}
\end{equation}
The result for $X$ is still negative definite, but the space derivatives dominate,
\begin{equation}
R(x) = \frac{\theta(\ell \!-\! \Vert \vec{x} \Vert)}{\ell^2} \Longrightarrow g^{\mu\nu}
\partial_{\mu} X \partial_{\nu} X = \Bigl( \frac{\partial X}{\partial r} \Bigr)^2 = 
\frac{r^2}{9 \ell^4} \theta(\ell \!-\! r) + \frac{\ell^2}{9 r^4} \theta(r \!-\! \ell) 
\; . \label{Xprimeflat}
\end{equation}
That reverses the sign of $Y$ from the cosmological case (\ref{flatXY}),
\begin{equation}
R(x) = \frac{\theta(\ell \!-\! \Vert \vec{x} \Vert)}{\ell^2} \Longrightarrow 
Y(x) = -\frac1{18} \Bigl[ \frac32 \!-\! \frac{r^4}{10 \ell^4}\Bigr] \theta(\ell \!-\! r)
-\frac1{18} \Bigl[ \frac{12 \ell}{5 r} \!-\! \frac{\ell^2}{r^2} \Bigr] \theta(r \!-\! \ell)
\; . \label{Yflat}
\end{equation}

The arguments we have given for $Y < 0$ are based on ignoring time derivatives of $X[g]$,
and should apply to systems which are strongly gravitationally bound. This should 
certainly be valid for the solar system, and even for the formation of large scale 
structure because peculiar velocities are expected to be small. However, around denser 
and/or warmer objects it may not be valid to ignore time derivatives of $X$, and this 
may modify the gravitational forces in these systems. This may perhaps turn out to be a 
critical test of our model.

\subsection{Unwanted Homogeneous Solutions}

For us the key insight of \cite{Belgacem:2018wtb} was that $X[g]$ is negative 
definite. However, it is worth commenting on the {\it additional} argument by
those authors that matching with the cosmological solution results in a small 
time-dependent homogeneous contribution to $X[g]$ inside gravitationally bound 
systems, extending over Megaparsec distances and even into strongly bound regions 
such as the solar system. This would induce a small time dependence in the 
effective Newton constant, which violates the constraints from lunar laser ranging. 

The effective Newton constant in our model is $G \times [1 + U + f(Y)]$ and it is 
obvious that the same argument does not necessarily apply to either $Y[g]$ or $U[g]$ 
because they are sourced differently from $X[g]$. On a deeper level, we question the 
plausibility of time-dependent homogeneous solutions carrying cosmological time 
dependence deep inside gravitationally bound systems even for $X[g]$. In general 
relativity the various gravitational fields also possess time-dependent homogeneous
solutions, yet there is no leakage of cosmological time dependence inside strongly
gravitationally bound systems. The authors of \cite{Belgacem:2018wtb} point out that 
they solved the scalar problem in the background of precisely such gravitational 
fields and still found time dependent solutions, but that is not realistic. The 
{\it mechanism} through which the gravitational fields of general relativity 
manage to avoid exciting time dependent homogeneous solutions inside bound systems
is feedback at the time the structure forms. Structures will form {\it differently}
in our model than in general relativity, and this will lead to feedback which
involves the auxiliary scalars as well as the gravitational fields. We think it
likely that this feedback will prevent the excitation of unwanted homogeneous
solutions for our model in the same way as it does in general relativity. However, 
this potential problem clearly deserves further study.

\subsection{Connection to Fundamentals}

We do not believe that nonlocality is fundamental; it is rather a conjecture for the 
most cosmologically significant part of the quantum gravitational effective action. 
The underlying idea \cite{Tsamis:1996qq} is that the problem of the cosmological 
constant \cite{Weinberg:1988cp,Carroll:2000fy} may have no resolution: general relativity 
really does have a large, positive cosmological constant, and this is what started 
primordial inflation. However, accelerated expansion led to the production of a vast 
ensemble of infrared gravitons \cite{Grishchuk:1974ny}, and the self-gravitation between 
these gravitons grew without bound as more and more came into causal contact. 
This self-gravitation provides a sort of quantum gravitational friction which slows 
inflation by an amount that eventually becomes nonperturbatively large. No one has yet 
devised a way of passing beyond perturbation theory to derive the result but the natural 
supposition is that this quantum gravitational effect eventually screened the large bare 
cosmological constant and brought inflation to a close.

Because what is being cancelled is a {\it constant}, whereas the screening 
mechanism is {\it dynamical}, depending on how many gravitons can see one another as
the past light-cone opens up, it is obvious that the persistence of perfect screening 
can only occur in one geometry. We believe this ``perfect'' geometry is radiation 
domination, and that the transition to matter domination disrupts perfect screening,
after which a small fraction of the original large bare cosmological constant peeks out 
from under the blanket of infrared gravitons which had previously completely screened it.

The key nonlocal ingredient in our model is the inverse scalar d`Alembertian 
$\square^{-1}$ which can be roughly motivated \cite{Tsamis:1997rk,Romania:2012av}
by the secular growth factors that arise in explicit loop corrections to gravitational 
radiation \cite{Tsamis:1996qk,Mora:2013ypa} and to gravitational forces 
\cite{Park:2011ww,Park:2015kua} on de Sitter background. For the rest, the Ricci scalar
is the simplest curvature scalar upon which it might act, and the combination in 
$Y[g](x)$ seems to be the simplest form which both matches the perturbative secular 
growth on de Sitter and also changes sign inside strongly gravitationally bound systems. 
At this stage there is of course no way to derive the nonlocal distortion function 
$f(Y)$, but simply accepting the model as a residual effect from the gravitational 
screening of inflationary gravitons does motivate two of its features which would 
otherwise seem parachuted in:
\begin{itemize}
\item{There is an initial value surface upon which the initial conditions of 
$\square^{-1}$ can be defined; and}
\item{There are modifications of gravity on large, but not small, distances without 
fine tuning or an explicit $\Lambda$}.
\end{itemize} 

\section{Enforcing the $\Lambda$CDM Expansion History}

In this section, we solve for the distortion function $f(Y)$
which supports the $\Lambda$CDM expansion history without dark energy. We begin by
specializing the model to cosmology, then describe the procedure for numerically
determining the required $f(Y)$. The section closes with a very simple and accurate
exponential fit to this function.

\subsection{The Cosmological Sector}

Cosmology's geometry is well described by a scale factor $a(t)$,
\begin{equation}
ds^2 = -dt^2 + a^2(t) d\vec{x} \!\cdot\! d\vec{x} \; , \label{FRW}
\end{equation}
whose expansion is quantified by the Hubble and first slow roll parameters,
\begin{equation}
H \equiv \frac{\dot{a}}{a} \qquad , \qquad \epsilon \equiv -\frac{\dot{H}}{H^2} \; .
\label{Heps}
\end{equation}
In this geometry the nonzero covariant derivative operators become,
\begin{equation}
\square \longrightarrow -\Bigl(\frac{d}{d t}\Bigr)^2 - 3 H \frac{d}{d t} \;\; , \;\;
D_0 D_0 \longrightarrow \Bigl( \frac{d}{d t}\Bigr)^2 \;\; , \;\; D_i D_j 
\longrightarrow g_{ij} H \frac{d}{d t} \; . \label{covders}
\end{equation}
The time-time component of the gravitational field equations (\ref{metriceqn}) is,
\begin{equation}
3 H \Bigl(\frac{d}{d t} + H\Bigr) \Bigl(1 + U + f(Y)\Bigr) + \frac12 \Bigl(
\dot{X} \dot{U} + \dot{Y} \dot{V} + \dot{X}^2\Bigr) = 8\pi G \rho \; , 
\label{Fried1}
\end{equation}
where $\rho$ is the energy density without dark energy. The space-space component 
is $g_{ij}$ times,
\begin{equation}
- \Bigl(\frac{d^2}{d t^2} + 2 H \frac{d}{d t} + 2 \dot{H} + 3 H^2\Bigr) \Bigl(1 + 
U + f(Y)\Bigr) + \frac12 \Bigl(\dot{X} \dot{U} + \dot{Y} \dot{V} + \dot{X}^2\Bigr) 
= 8\pi G p \; , \label{Freid2}
\end{equation}
where $p$ is the pressure, also without dark energy.

\begin{figure}[H]
\includegraphics[width=6.0cm,height=4.0cm]{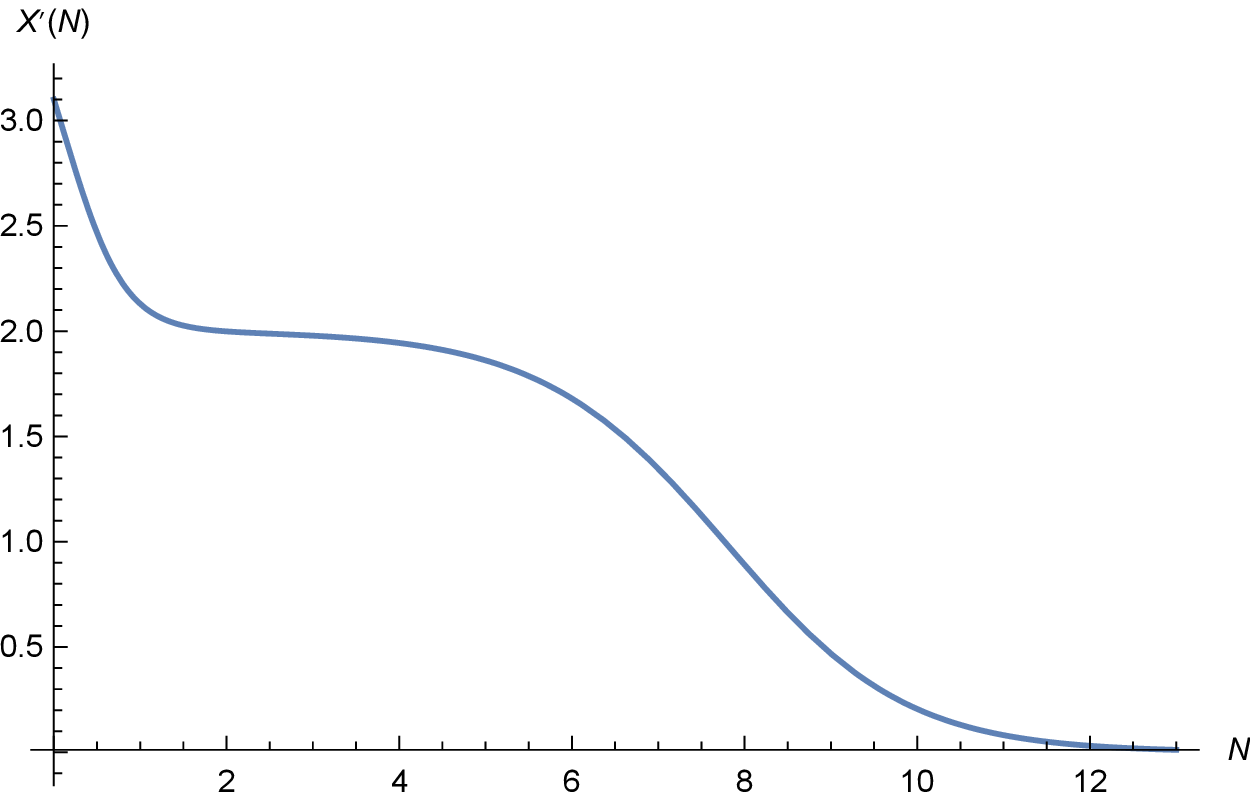}
\hspace{1cm}
\includegraphics[width=6.0cm,height=4.0cm]{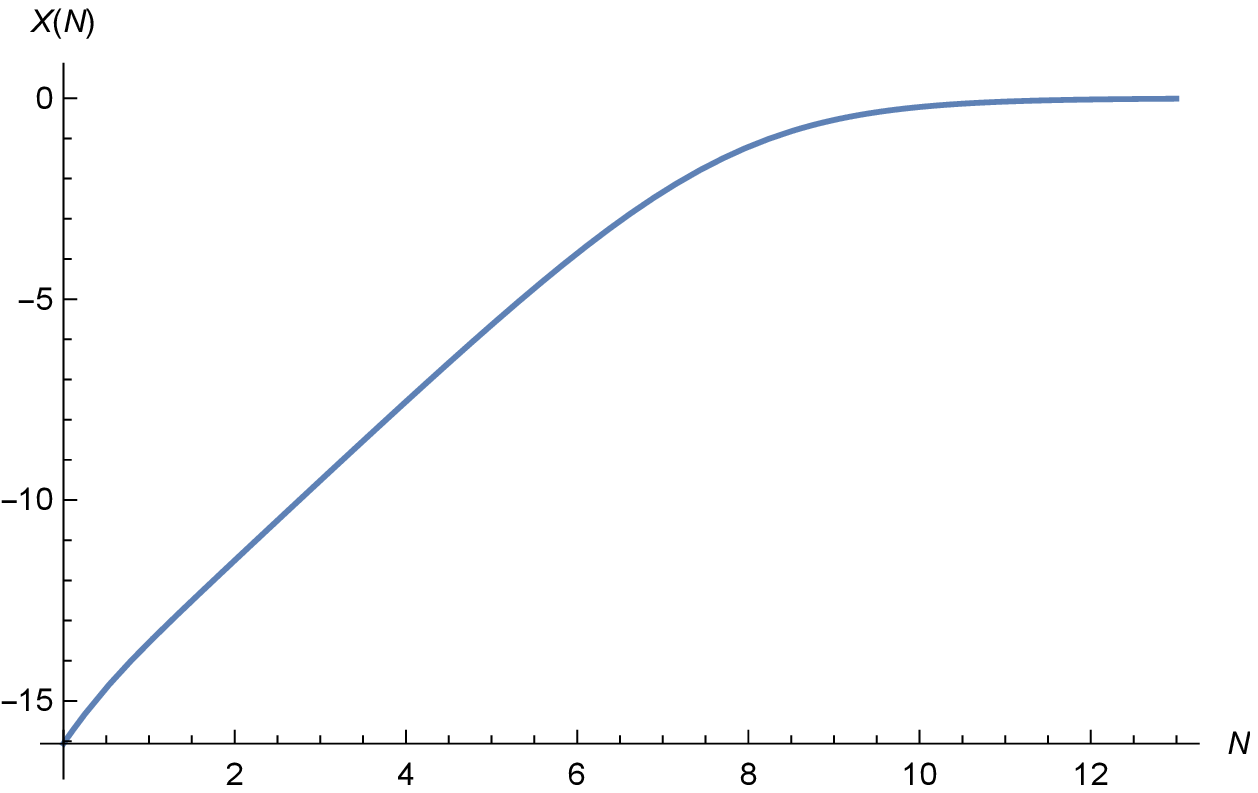}
\includegraphics[width=6.0cm,height=4.0cm]{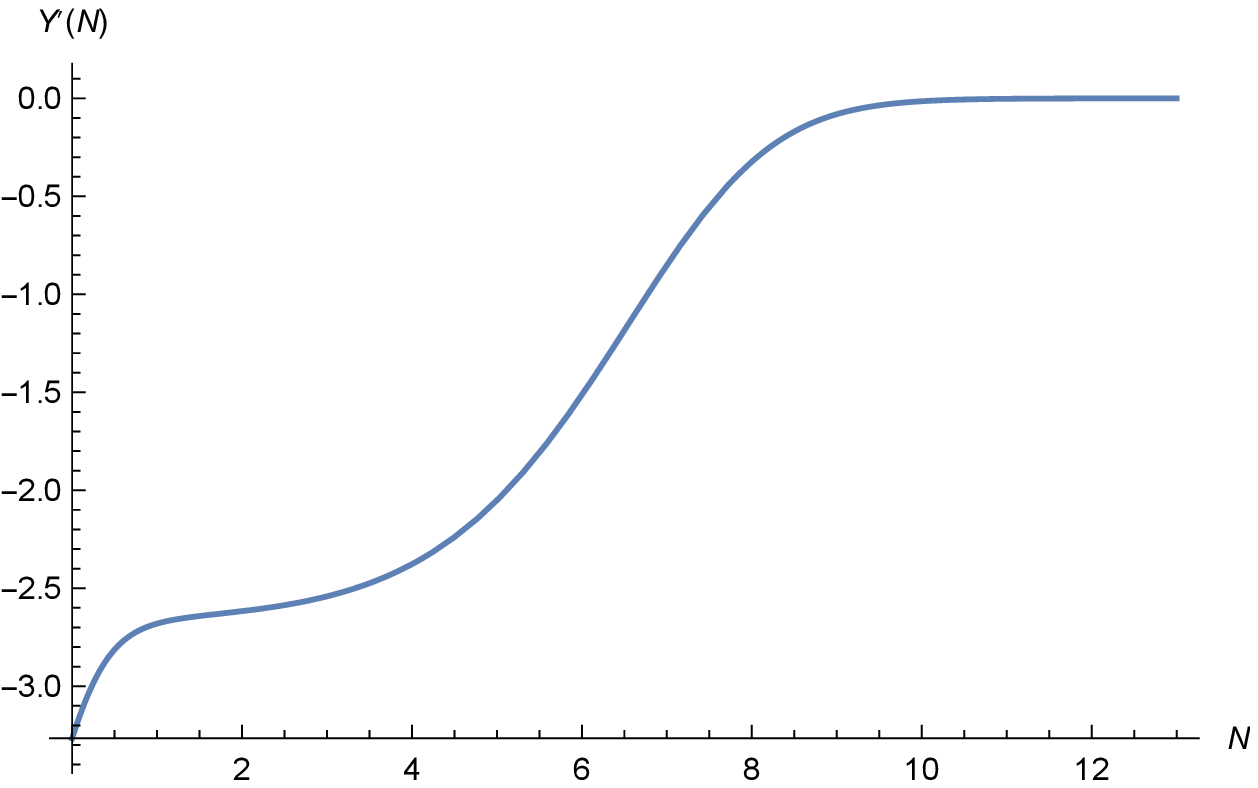}
\hspace{1cm}
\includegraphics[width=6.0cm,height=4.0cm]{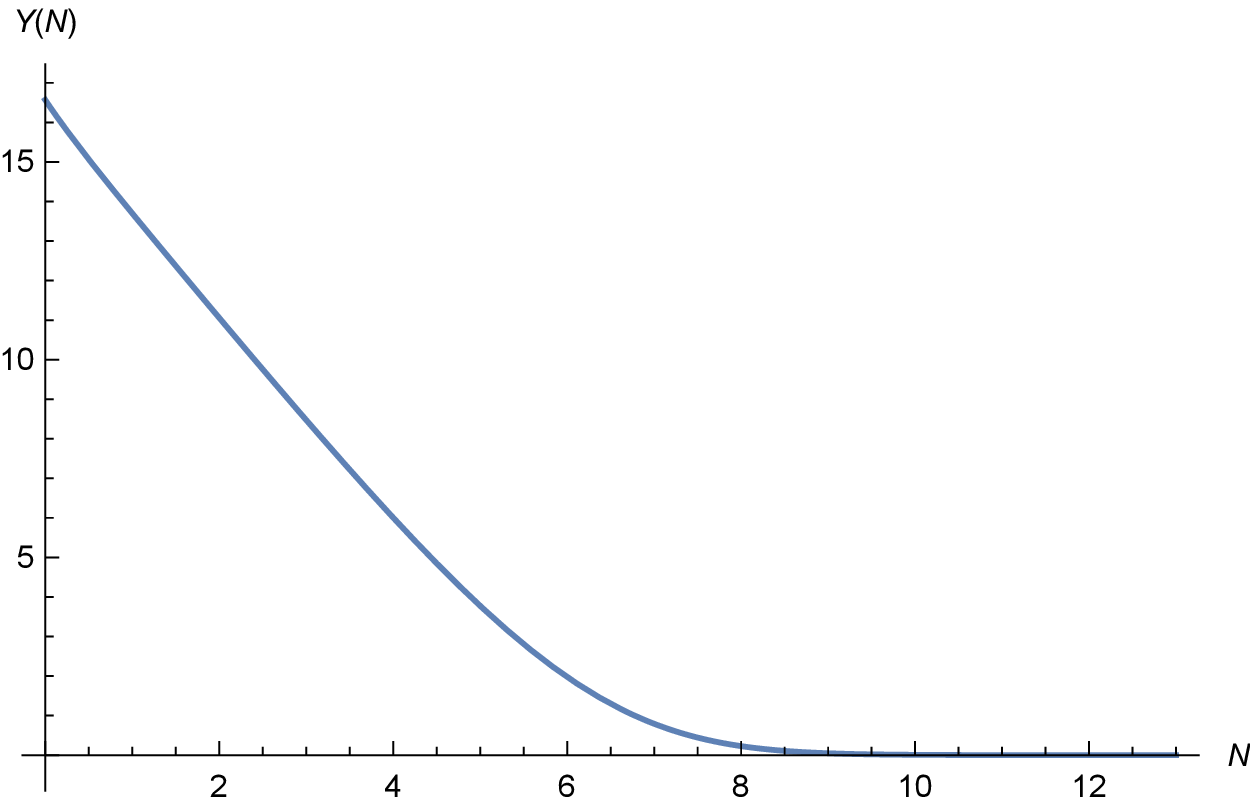}
\caption{The left hand graphs show numerical simulations of $X'(N)$ and $Y'(N)$
as defined by (\ref{Xprime}-\ref{Yprime}). The right hand graphs give their 
integrals.}
\label{XYofN}
\end{figure}

The best time variable is $N \equiv \ln(\frac{a_0}{a})$, the number of e-foldings 
until the present. The various differentials and derivatives then simplify,
\begin{equation}
dN = -H dt \quad , \quad \frac{d}{d t} = -H \frac{d}{d N} \quad , \quad
\frac{d^2}{d t^2} = H^2 \Bigl( \frac{d^2}{d N^2} + \epsilon \frac{d}{d N} \Bigr) 
\; . \label{timetoN}
\end{equation}
We seek to determine the function $f(Y)$ to enforce the $\Lambda$CDM expansion
history without a cosmological constant. This means the Hubble parameter, energy
density and pressure take the forms,
\begin{eqnarray}
H^2 & = & H_0^2 \Bigl( \Omega_r e^{4N} + \Omega_m e^{3N} + \Omega_{\Lambda}\Bigr)
\equiv H_0^2 \!\times\! \widetilde{H}^2 \; , \label{HLCDM} \\
8 \pi G \rho & = & 3 H_0^2 \Bigl( \Omega_r e^{4 N} + \Omega_m e^{3 N}\Bigr) 
\equiv H_0^2 \!\times\! \widetilde{\rho} \; , \label{rhoLCDM} \\
8 \pi G p & = & 3 H_0^2 \times \frac13 \Omega_r e^{4 N} \equiv H_0^2 \!\times\!
\widetilde{p} \; , \label{pLCDM}
\end{eqnarray}
where $H_0$ is the current value of the Hubble parameter and $\Omega_r$, 
$\Omega_m$ and $\Omega_{\Lambda} = 1 - \Omega_r - \Omega_m$ are the $\Lambda$CDM
fractions of the energy density in radiation, matter and vacuum energy.\footnote{
We use $\Omega_{m}/\Omega_{r} \equiv 1 + z_{eq} \simeq 3403$, $\Omega_{m} \simeq 
0.3153$ and $\Omega_{\Lambda} \simeq 0.6847$ \cite{Aghanim:2018eyx}.}
In this notation the scalar equations are,
\begin{eqnarray}
\Bigl[ \widetilde{H} e^{-3N} X'\Bigr]' = -12 \Bigl(1 \!-\! \frac12 \epsilon\Bigr) 
\widetilde{H} e^{-3N} \; &\!\! , \!\!& \; \Bigl[ \widetilde{H} e^{-3N} Y'\Bigr]' 
= \widetilde{H} e^{-3N} {X'}^2 \; , \qquad \label{XYofNeqn} \\
\Bigl[ \widetilde{H} e^{-3N} V'\Bigr]' = -12 \Bigl(1 \!-\! \frac12 \epsilon\Bigr) 
\widetilde{H} e^{-3N} f'(Y) \; &\!\! , \!\!& \; U' = -2 X' V \; , \label{UVofN}
\end{eqnarray} 
where a prime denotes differentiation with respect to the natural argument --- 
$Y$ for $f(Y)$ and $N$ for $\widetilde{H}(N)$, $X(N)$, $Y(N)$, $U(N)$ and $V(N)$. 
Note that equations (\ref{XYofNeqn}) give explicit integral expressions for $X'(N)$
and $Y'(N)$,
\begin{eqnarray}
X'(N) & = & \frac{e^{3N}}{\widetilde{H}(N)} \! \int_{N}^{\infty} \!\! dN' \,
\frac{e^{-3 N'}}{\widetilde{H}(N')} \Bigl[3 \Omega_{m} e^{3 N'} + 12 
\Omega_{\Lambda}\Bigr] \; , \label{Xprime} \\
Y'(N) & = & -\frac{e^{3N}}{\widetilde{H}(N)} \! \int_{N}^{\infty} \!\! dN' \,
e^{-3 N'} \widetilde{H}(N') \Bigl[X'(N')\Bigr]^2 \; .
\label{Yprime}
\end{eqnarray}
Figure \ref{XYofN} shows $X(N)$ and $Y(N)$ and their derivatives.
\noindent Since $X'(N)$ and $Y'(N)$ have definite signs here, both $X$ and $Y$ 
are monotonic, hence invertible.

\subsection{The Reconstruction Procedure}

The two gravitational field equations are,
\begin{eqnarray}
-3 (\partial_N \!-\! 1) \Bigl[ U \!+\! f(Y)\Bigr] + \frac12 \Bigl[ X' U' \!+\! 
Y' V' \!+\! {X'}^2 \Bigr] & \!\!\!=\!\!\! & -\frac{3 \Omega_{\Lambda}}{
\widetilde{H}^2} \; , \qquad \label{F1ofN} \\
-\Bigl( \partial_N^2 \!-\! (2 \!-\! \epsilon) \partial_N \!+\! 3 \!-\! 2\epsilon 
\Bigr) \Bigl[U \!+\! f(Y)\Bigr] \!+\! \frac12 \Bigl[ X' U' \!+\! Y' V' \!+\! 
{X'}^2\Bigr] & \!\!\!=\!\!\! & \frac{3 \Omega_{\Lambda}}{\widetilde{H}^2} \; . 
\label{F2ofN}
\end{eqnarray}
As for the original model \cite{Deffayet:2009ca}, the first step in 
constructing a nonlocal distortion function which supports the $\Lambda$CDM 
expansion history is to take the difference of (\ref{F1ofN}) and (\ref{F2ofN}),
\begin{equation}
(\partial_N - 3 + \epsilon) (\partial_N - 2) \Bigl[ U + f(Y)\Bigr] = 
-\frac{6 \Omega_{\Lambda}}{\widetilde{H}^2} \; . \label{step1}
\end{equation}
This can be integrated to give exactly the same result as for the original model
\cite{Deffayet:2009ca},
\begin{equation}
U + f(Y) = -6\Omega_{\Lambda} e^{2 N} \!\! \int_{N}^{\infty} \!\! dN' \,
\frac{e^{N'}}{\widetilde{H}(N')} \! \int_{N'}^{\infty} \!\! dN'' \,
\frac{e^{-3 N''}}{\widetilde{H}(N'')} \equiv g(N) \; . \label{step2}
\end{equation}

The next step is to derive a differential equation for the function,
\begin{equation}
G(N) \equiv \frac{Y'(N)}{X'(N)} f'\Bigl(Y(N)\Bigr) - \frac{g'(N)}{X'(N)} \; . 
\label{Gdef} 
\end{equation}
Differentiating relation (\ref{step2}), using (\ref{UVofN}) and dividing by
$X'(N)$ gives,
\begin{equation}
-2 V(N) + G(N) = 0 \; . \label{step3}
\end{equation}
Acting $\partial_N^2 - (3 - \epsilon) \partial_N$ on (\ref{step3}) and 
using relation (\ref{UVofN}) produces,
\begin{equation}
(\partial_N - 3 + \epsilon) \partial_N G 
+ 24 \Bigl(1 - \frac12 \epsilon\Bigr) \frac{X'}{Y'} G
= -24 \Bigl(1 - \frac12 \epsilon\Bigr) \frac{g'}{Y'} \; . \label{step4}
\end{equation}
The procedure from this point is clear: we numerically solve (\ref{step4})
for $G(N)$, extract $\partial _N f(Y) = Y' \times f'(Y)$ using relation 
(\ref{Gdef}), numerically integrate to recover $f(Y)$ as a function of $N$,
and finally exploit the one-to-one relation between $Y$ and $N$ to 
numerically express $f(Y)$ as a function of $Y$. 

\subsection{Solution for $f(Y)$}

The initial conditions at large $N$ follow from exact results, derived in 
Appendix A, by retaining only the leading dependence on $\Omega_{\Lambda}$.
Because $\Omega_{\Lambda}$ is irrelevant until late times, expressions 
(\ref{largeNXp}-\ref{largeNgp}) are accurate to three digits for $N > 2$. The 
functions we need for equation (\ref{step4}) can be usefully expanded in powers 
of the variable $y \equiv (1 + z_{eq}) e^{-N}$,
\begin{eqnarray}
\epsilon & \longrightarrow & +2 -\frac12 y + \frac12 y^2 - \frac12 y^3 + 
O(y^4) \; , \label{earlyepsilon} \\
X' & \longrightarrow & +\frac32 y - \frac54 y^2 + \frac{35}{32} y^3 - 
\frac{63}{64} y^4 + O(y^5) \; , \label{earlyXprime} \\
Y' & \longrightarrow & -\frac34 y^2 + \frac{33}{32} y^3 - \frac{367}{320} y^4 +
\frac{4577}{3840} y^5 + O(y^6) \; , \label{earlyYprime} \\
g' & \longrightarrow & \frac{\Omega_{\Lambda} \Omega_{r}^3}{\Omega_{r}^4}
\Biggl\{ \frac45 y^4 - \frac{11}{14} y^5 + \frac{429}{560} y^6 - \frac{142}{192}
y^7 + O(y^8) \Biggr\} \; . \label{earlygprime}
\end{eqnarray}
Because $1 + z_{eq} \simeq 3403 \simeq \exp[+8.132]$ is so large, these 
expansions are only accurate for $N > 10$. Employing the expansions
(\ref{earlyepsilon}-\ref{earlygprime}) allows us to factorize the large $N$
limiting form of the differential operator in equation (\ref{step4}),
\begin{equation}
F_1(y) \frac{d}{d N} \Biggl\{ F_2(y) \frac{d}{d N} \Bigl[ F_3(y) G\Bigr] \Biggr\}
= F_4(y) \; , \label{largeNeqn}
\end{equation}
where the four factors are,
\begin{eqnarray}
F_1(y) & = & \frac1{y^4} \Biggl[1 \!+\! \frac38 y \!-\! \frac{13}{960} y^2 \!-\! 
\frac{13}{4608} y^3 + \dots \Biggr] \; , \label{F1} \\
F_2(y) & = & y^7 \Biggl[1 \!-\! \frac54 y \!+\! \frac{1151}{960} y^2 - 
\frac{6071}{5760} y^3 + \dots \Biggr] \; , \label{F2} \\
F_3(y) & = & \frac1{y^3} \Biggl[1 \!+\! \frac78 y \!+\! \frac{47}{960} y^2 
\!+\! \frac{137}{23040} y^3 + \dots \Biggr] \; , \label{F3} \\
F_4(y) & = & \frac{\Omega_{\Lambda} \Omega_{r}^3}{\Omega_{m}^4} 
\Biggl\{\frac{32}{5} y^3 \!-\! \frac{136}{35} y^4 \!+\! \frac{3869}{1050} y^5 
\!-\! \frac{2587}{720} y^6 + \dots \Biggr\} . \label{F4}
\end{eqnarray}
Expression (\ref{largeNeqn}) is a second order differential equation and
possesses two homogeneous solutions. However, only one of these falls off for 
large $N$,
\begin{equation}
G_{\rm h}(N) = \frac1{F_3(y)} = y^3 - \frac78 y^4 + \frac{43}{60} y^5 - 
\frac{85}{144} y^6 + O(y^7) \; . \label{Ghomo}
\end{equation}
The large $N$ limiting form of $G(N)$ can be inferred from (\ref{largeNeqn}),
\begin{equation}
G(N) \longrightarrow \frac{\Omega_{\Lambda} \Omega_{r}^3}{\Omega_{m}^4} 
\Biggl\{ \frac{32}{35} G_{\rm h}(N) \ln(y) + \frac{5}{14} y^4 - 
\frac{1247}{4200} y^5 + \frac{71117}{302400} y^6 + O(y^7) \Biggr\} . 
\label{largeNG}
\end{equation}
This provides the initial conditions to evolve (\ref{step4}) from finite $N$.

\begin{figure}[H]
\includegraphics[width=6.0cm,height=4.0cm]{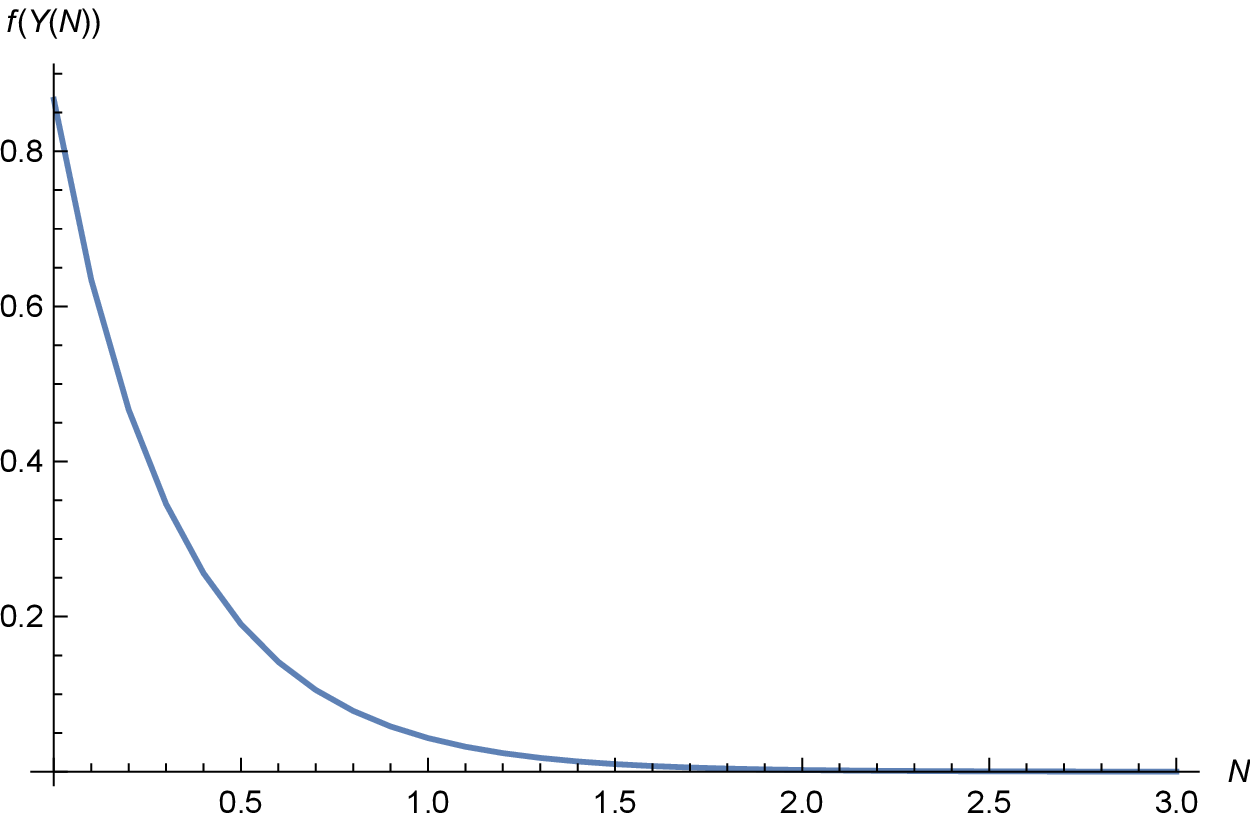}
\hspace{1cm}
\includegraphics[width=6.0cm,height=4.0cm]{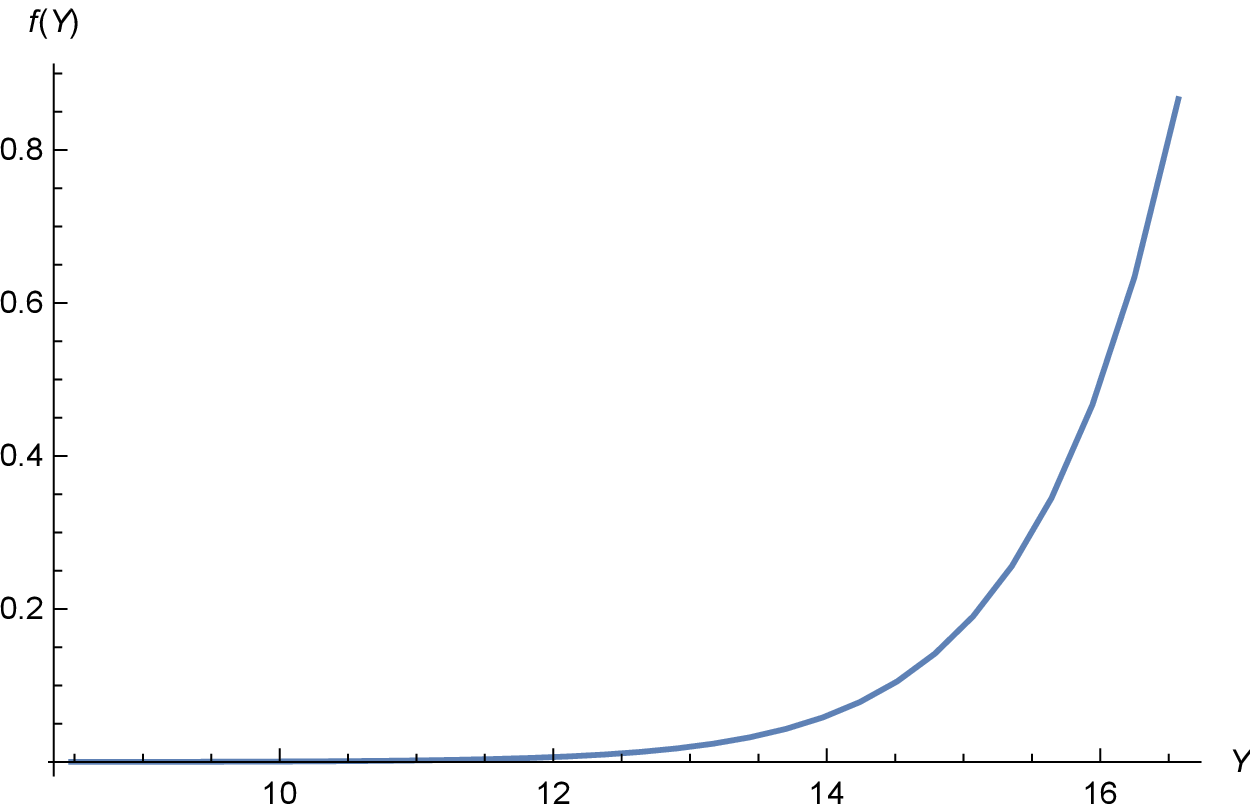}
\caption{The left hand graph shows a numerical simulation of $f(Y)$ as a 
function of the evolution variable $N$. The right hand graph also gives $f(Y)$, 
but now as a function of its natural argument $Y$.}
\label{fofY}
\end{figure}

The expansion (\ref{largeNG}) fixes the small $Y$ behavior of $f(Y)$,
\begin{equation}
f(Y) = \frac{\Omega_{\Lambda} \Omega_{r}^3}{\Omega_{m}^4} 
\Biggl\{ -\frac{128}{105} Y^2 \ln(Y) + O(Y^2) \Biggr\} . \label{smallY}
\end{equation}
This means that making $f(Y)$ vanish for all $Y < 0$ leads to no discontinuity
in either $f(Y)$ or $f'(Y)$ at $Y = 0$. Numerical evolution gives the result
for general $Y > 0$, which is depicted in Fig.~\ref{fofY}.
\begin{figure}[H]
\includegraphics[width=6.0cm,height=4.0cm]{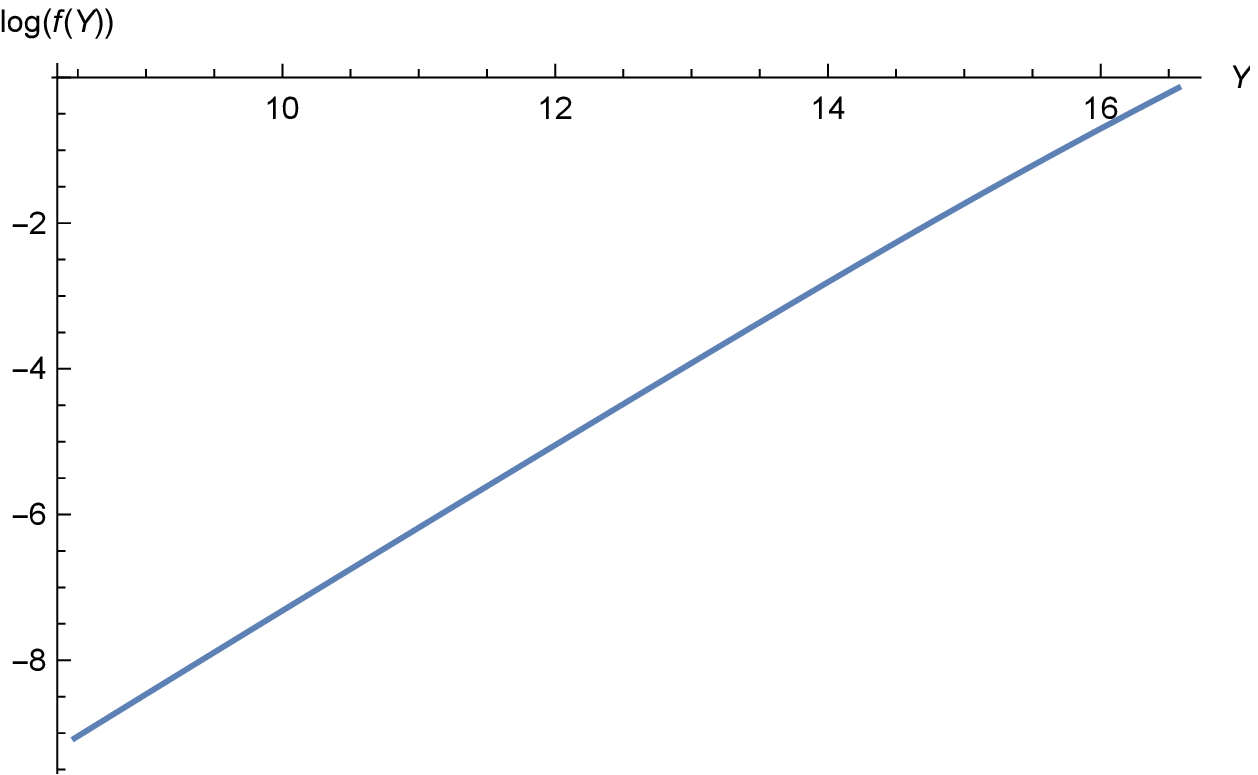}
\hspace{1cm}
\includegraphics[width=6.0cm,height=4.0cm]{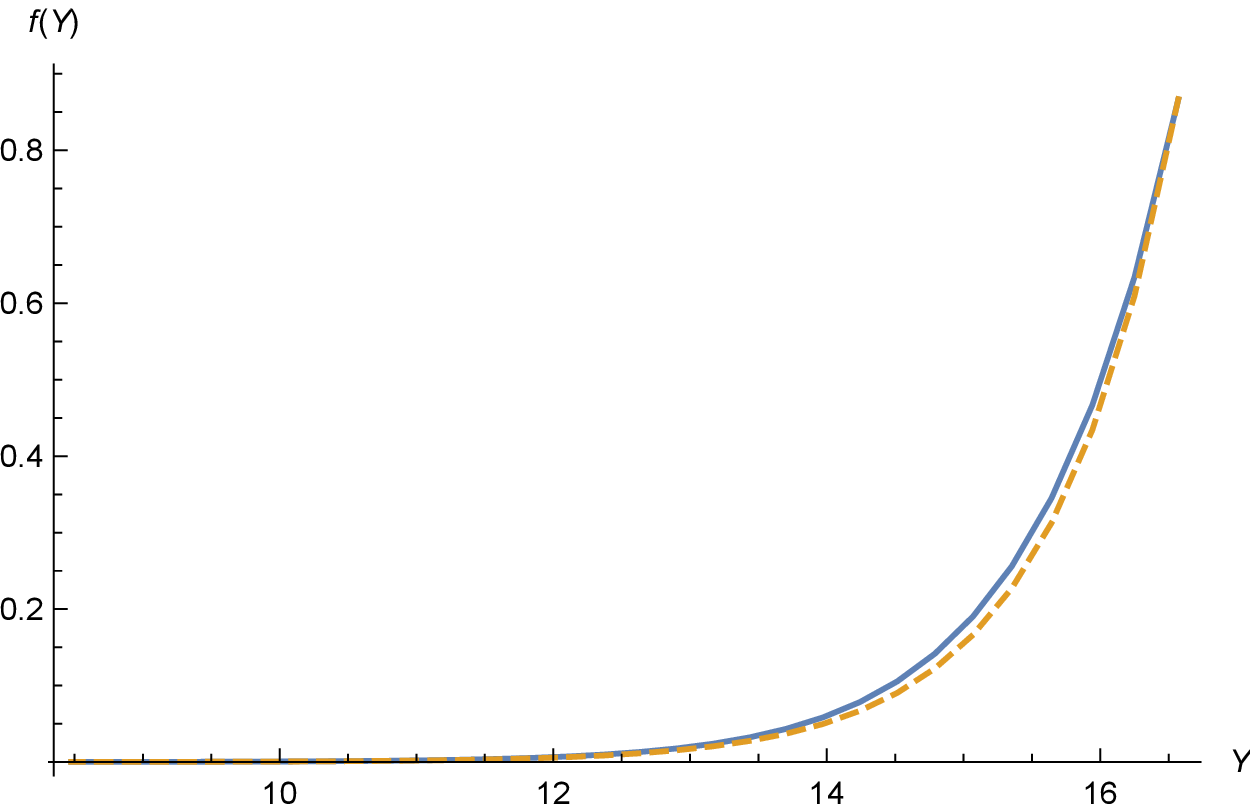}
\caption{The left hand graph shows that $\ln[f(Y)]$ is nearly a straight line. 
The right hand graph compares the full numerical determination of $f(Y)$ (in solid, 
blue) to the resulting exponential fit (\ref{fit}) (in dashed, yellow).}
\label{logfit}
\end{figure}
\noindent Figure~\ref{logfit} shows that $f(Y)$ is well fit by the strikingly 
simple form, duly matched to (\ref{smallY}) at small $Y$,\footnote{Note that this 
form only pertains for positive $Y$ somewhat greater than zero. We still require 
$f(Y)$ to vanish as in (\ref{smallY}) as $Y \rightarrow 0^{+}$, and to vanish for 
all $Y \leq 0$.}
\begin{equation}
f(Y) \simeq \exp\Bigl[ 1.1 \Bigl(Y - 16.7\Bigr) \Bigr] \; . \label{fit}
\end{equation}
 
\section{Discussion}

We have presented a simple variant of our original model \cite{Deser:2007jk}
to explain the current phase of cosmic acceleration without dark energy.
Like its ancestor, the new model is based on augmenting the Hilbert action by the
addition of $R$ times a function of a dimensionless, nonlocal scalar;
only the scalar has changed from $X[g] = \square^{-1} R$ to $Y[g] = \square^{-1} 
g^{\mu\nu} \partial_{\mu} X \partial_{\nu} X$. Both $X[g]$ and $Y[g]$ are 
quiescent during radiation domination, and thereafter only grow logarithmically, 
which provides a natural explanation for why the onset of acceleration is delayed
to late in cosmic history. Both scalars also vanish for gravitational radiation
which means that they do not affect the --- tightly constrained --- propagation 
velocity \cite{Boran:2017rdn}.

Because section 3 employed the parameter $\Omega_{\Lambda}$ to make our model
reproduce the $\Lambda$CDM expansion history, one might question our claim of
no fine tuning. However, it is best to view fine tuning from the perspective of 
how precisely the parameters of the Lagrangian must be adjusted to explain late
time acceleration. Our model (\ref{Lnonlocal}) amounts to replacing the standard 
$R$ term of general relativity by $R + R \times f(Y)$, where $Y[g]$ obeys 
(\ref{XYdef}). In this model nothing needs to be done to delay the onset of 
acceleration until very late in cosmological history; that happens naturally 
because $Y[g]$ is sourced by $R$, which vanishes during radiation domination, 
and because $Y[g]$ only grows logarithmically after the transition to matter 
domination. Nor are there any new scales; the function $f(Y)$ is dimensionless, 
as is $Y[g]$ itself. As long as $f(Y)$ grows with $Y$ there will be a phase of 
late time acceleration. Just how little the old model \cite{Deser:2007jk} 
changes with variations of the nonlocal distortion function has already been 
explored \cite{Park:2016jym} and would not differ in the new model. Contrast 
this with the two local alternatives of general relativity with a cosmological 
constant or a scalar quintessence model,
\begin{eqnarray}
R & \longrightarrow & R - 2 \Lambda \label{GRLambda} \; , \\
R & \longrightarrow & R - \frac12 \partial_{\mu} \varphi \partial_{\nu} \varphi
g^{\mu\nu} - V(\varphi) \; . \label{Squint}
\end{eqnarray}
In the first case (\ref{GRLambda}) the dimensionful parameter $\Lambda$ must be
fine-tuned to make the dimensionless product $G \Lambda \simeq 10^{-122}$ vanish
to 122 decimal places! A similar amount of fine tuning must be imposed on the
potential $V(\varphi)$ of quintessence models (\ref{Squint}). From this 
perspective, our model is indeed a non-fine-tuned one!

\begin{figure}[H]
\includegraphics[width=6.0cm,height=4.0cm]{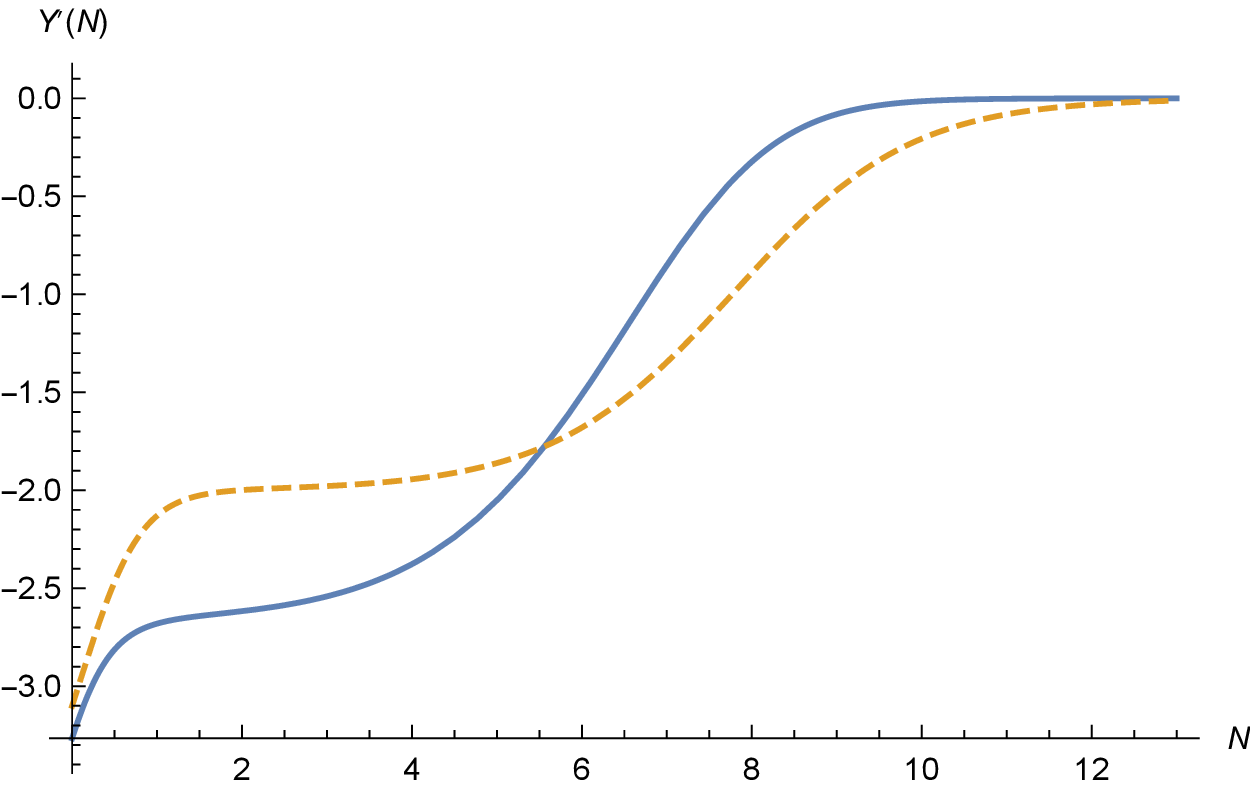}
\hspace{1cm}
\includegraphics[width=6.0cm,height=4.0cm]{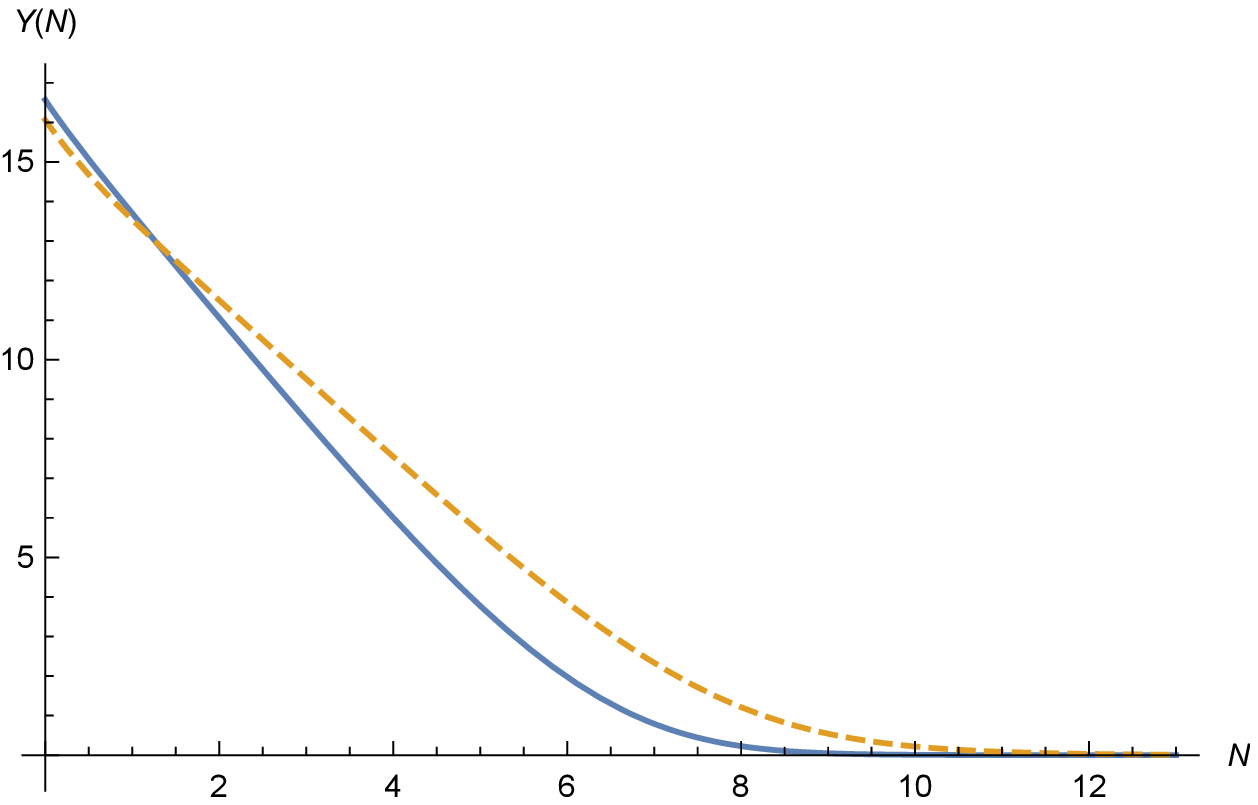}
\caption{The left hand graph compares $Y'(N)$ (solid blue) with $-X'(N)$ 
(dashed yellow). The right hand graph compares $Y(N)$ (solid blue) with $-X(N)$ 
(dashed yellow).}
\label{YXcomp}
\end{figure}

Fig.~\ref{YXcomp} shows that $Y$ is close to $-X$ for cosmology.
\noindent  This made it simple to determine the nonlocal distortion
function $f(Y)$ numerically in order to reproduce exactly the $\Lambda$CDM 
expansion history without dark energy. That was done in section 3, with results 
shown on Fig.~\ref{fofY}. An unexpected consequence was the simple exponential 
approximation (\ref{fit}) for $f(Y)$, whose accuracy can be seen from 
Fig.~\ref{logfit}. 

The new model differs from the original one in that $Y$ (unlike $X$) changes sign
from cosmology (with $Y > 0$) to strongly gravitationally bound systems (with 
$Y < 0$). Because cosmology only fixes $f(Y)$ for $Y \geq 0$, with $f(0) = 0$, 
simply assuming $f(Y) = 0$ for $Y < 0$ protects the model from changing the 
heavily constrained physics of the solar system. The huge advantages of this 
model can be seen by comparison with $F(R)$ theories of gravity, which must invoke 
ever more exotic physics such as the chameleon mechanism \cite{Khoury:2003aq}
to evade solar system constraints. Note also that the only stable choice of
$F(R)$ which exactly reproduces the $\Lambda$CDM expansion history is $F(R) = 
R - 2\Lambda$ \cite{Dunsby:2010wg}.

Now that the nonlocal distortion function $f(Y)$ has been fixed the model is
complete. Because $f(Y)$ has been chosen to exactly reproduce the $\Lambda$CDM 
expansion history, with no changes inside gravitationally bound systems, tests
of the model must come from its predictions for cosmological perturbations and
the growth of structures. Stability is another important constraint to study.

Finally, we return to the presumed local sources of our model, the gravitons
of primordial inflation. Their loop effects can grow non-perturbatively 
strongly during the primordial inflation era \cite{Mora:2013ypa}, and 
$\square^{-1}$ does correctly capture this growth on de Sitter background,
but it is clearly a major unsolved problem to follow their temporal effects 
in any detail. We have no explanation, other than simplicity and 
dimensionlessness, for the combination $X = \square^{-1} R$, nor can we 
justify the appearance of $Y$. It might, however, be worth noting that nonlocal 
realizations of MOdified Newtonian Dynamics (MOND) \cite{Milgrom:1983ca,
Milgrom:1983pn,Milgrom:1983zz} involve a similar nonlocal scalar 
\cite{Deffayet:2011sk,Deffayet:2014lba}. This raises the hope that there is a 
master effective action describing the full range of cosmic history from the 
build-up of gravitational back-reaction during inflation, and giving rise to 
both the present model and to MOND as residual effects.

For now we can strictly only offer our phenomenological (but dimensionless) 
construction. Nevertheless, the presumed inflationary origin does provide two 
vital answers that otherwise seem unnatural: the existence of an initial value 
surface from which one may launch the initial conditions defining our inverse  
differential operators, i.e., the propagators, {\it and} why the 
corrections only modify classical general relativity on cosmological, rather 
than on the smaller (bound matter) scales, where no ``improvement'' is needed!

\vskip 1cm

\centerline{\bf Acknowledgements}

We are grateful to M. Maggiore for calling our attention to the 
problem with the original model. SD's work was supported by the U.S. Department
of Energy, Office of Science, Office of High Energy Physics, under Award Number 
de-sc0011632. RPW was partially supported by NSF grants PHY-1506513 and PHY-1806218, 
and by the Institute for Fundamental Theory at the University of Florida.

\section{Appendix: Exact Expressions to Leading Order in $\Omega_{\Lambda}$}

Expressions for $X(N)$, $Y(N)$ and $g(N)$ simplify dramatically for
large $N$ when $\Omega_{\Lambda}$ becomes negligible relative to 
$\Omega_{m} e^{3 N}$ and $\Omega_{r} e^{4 N}$. Setting $\Omega_{\Lambda} = 0$
and making the change of variable $y \equiv \frac{\Omega_{m}}{\Omega_{r}} 
e^{-N}$ in expression (\ref{Xprime}) reduces $X'(N)$ to an elementary
function,
\begin{equation}
X'(N) \longrightarrow \frac{3}{y \sqrt{1 \!+\! y}} \! \int_{0}^{y} \!
\frac{y' dy'}{\sqrt{1 \!+\! y'}} = \frac{2 (z \!-\! 1) (z \!+\! 2)}{z
(z \!+\! 1)} \; , \label{largeNXp}
\end{equation}
where $z \equiv \sqrt{1 + y}$. Integrating expression (\ref{largeNXp}) gives,
\begin{equation}
X(N) \longrightarrow -2 \Bigl( \frac{z \!-\! 1}{z \!+\! 1}\Bigr) - 
4 \ln\Bigl(\frac12 z \!+\! \frac12\Bigr) \; . \label{largeNX}
\end{equation}

Setting $\Omega_{\Lambda} = 0$ in expression (\ref{Yprime}) similarly 
reduces $Y'(N)$ to an elementary function,
\begin{eqnarray}
Y'(N) & \longrightarrow & \frac{-1}{y \sqrt{1 \!+\! y}} \! \int_{0}^{y} \!\! 
dy' \, \sqrt{1 \!+\! y'} \, \Bigl[ X'(N') \Bigr]^2 \; , \\
& = & \frac{-8}{z (z \!+\! 1)} \Biggl[ \frac13 \Bigl( z^2 \!+\! z \!-\! 11
\Bigr) + \frac{2}{z \!+\! 1} + \frac{4 \ln(\frac12 z \!+\! \frac12)}{z \!-\! 1}
\Biggr] \; . \qquad \label{largeNYp} 
\end{eqnarray}
Integrating (\ref{largeNYp}) to get $Y(N)$ produces a dilogarithm in 
addition to elementary functions,
\begin{eqnarray}
Y(N) & \longrightarrow & \frac{8}{(z \!+\! 1)^2} - \frac{112}{3 (z \!+\! 1)} 
+ \Bigl[ \frac{16}{3} - \frac{32 z}{z^2 \!-\! 1} \Bigr] \ln\Bigl(\frac12 z 
\!+\! \frac12 \Bigr) + \frac{37}{24} \nonumber \\
& & \hspace{4cm} + 8 \ln^2\Bigl(\frac12 z \!+\! \frac12\Bigr) + 16
{\rm Li}_2\Bigl( \frac12 \!-\! \frac12 z\Bigr) \; , \label{largeNY} \qquad
\end{eqnarray}
where
\begin{equation}
{\rm Li}_2(x) \equiv -\int_{0}^{x} \!\! dt \, \frac{\ln(1 \!-\! t)}{t} \; .
\label{dilog}
\end{equation}

The function $g(N)$ actually vanishes with $\Omega_{\Lambda}$ so its large 
$N$ limit derives from preserving the initial factor in (\ref{step2}),
\begin{eqnarray}
g(N) & \longrightarrow & -\frac{6 \Omega_{\Lambda} \Omega_{r}^3}{
\Omega_{m}^4 y^2} \! \int_{0}^{y} \! \frac{dy'}{\sqrt{1 \!+\! y'}} \!
\int_{0}^{y'} \! dy'' \, \frac{ {y''}^4}{\sqrt{1 \!+\! y''}} \; , \\
& = & -\frac{\Omega_{\Lambda} \Omega_{r}^3}{\Omega_{m}^4} 
\frac{(z \!-\! 1)^4}{105} \Biggl[ 28 z^2 + 112 z + 156 + \frac{64}{z \!+\! 1}
+ \frac{32}{(z \!+\! 1)^2} \Biggr] \; . \label{largeNg} \qquad
\end{eqnarray}
Differentiating this gives,
\begin{equation}
g'(N) \longrightarrow \frac{\Omega_{\Lambda} \Omega_{r}^3}{\Omega_{m}^4}
\frac{(z \!-\! 1)^4}{105} \Biggl[ 84 z^2 + 336 z + 508 + \frac{352}{z} +
\frac{96}{z (z \!+\! 1)} + \frac{96}{z (z+1)^2} \Biggr] . \label{largeNgp}
\end{equation}
All of these expressions are accurate to  three digits for $N > 2$.

\end{document}